\documentclass{article}

\usepackage{amsfonts}
\usepackage{amsmath}
\usepackage{amsthm}
\usepackage{graphicx}

\newcommand{\R}{\mathbb R}
\newcommand{\N}{\mathbb N}
\newcommand{\n}{\underline{n}}

\newcommand{\norm}[1]{\left\Vert#1\right\Vert}

\newcommand{\eps}{\varepsilon}

\newcommand{\unit}{E}

\newcommand{\DW}{\mathrm{DW}}
\newcommand{\HK}{\mathrm{HK}}

\newtheorem{lemma}{Lemma}
\newtheorem{theorem}[lemma]{Theorem}
\newtheorem{proposition}[lemma]{Proposition}

\begin{document}
%\doi{}
% \issn{1563-5120}
% \issnp{1023-6198} %\jvol{00} \jnum{00} \jyear{2006} \jmonth{August}

% \markboth{Taylor \& Francis and A. Consultant}{\LaTeXe\ guide for
%   authors}

\title{Fixed points in models of continuous opinion dynamics under
  bounded confidence}

\author{Jan Lorenz\thanks{Email: post@janlo.de,
    present adress: ETH Z\"urich, Chair of Systems Design, Kreuzplatz~5,
    8032 Z\"urich, Switzerland}\\
  Universit\"at Bremen, Fachbereich Mathematik und Informatik\\
  Bibliothekstra\ss{}e, 28359 Bremen, Germany \\\vspace{6pt}
  %\received{v3.1 released August 2006}
 }
\date{October 6, 2007}

\maketitle

\begin{abstract}
  We present two models of continuous opinion dynamics under bounded
  confidence which are representable as nonnegative discrete dynamical
  systems, namely the Hegselmann-Krause model (Hegselmann and Krause,
  Journal of Artificial Societies and Social Simulation 5(3), 2002)
  and the Deffuant-Weisbuch model (Deffuant {\em et al}, Advances in Complex Systems,
  3, 2000). We fully characterize the set of fixed points for both
  models. They are identical. Further on, we present reformulations of
  both models on the more general level of densities of agents in the
  opinion space as interactive Markov chains. We also characterize the sets of fixed
  points as identical in both models. 

\bigskip

  \begin{center}
    agent-based models, density-based models, interactive Markov chain,
    discrete master equation
  \end{center}
\end{abstract}

\section{Introduction}

Consider a set of $n\in\N$ agents which hold continuous opinions.
`Continuous' means that the opinion is in its essence a vector of
$d\in\N$ real numbers. An example for a continuous opinion is a budget
plan proposal, where a fixed amount of money is distributed to $d$
departements. Other examples are prices for products or an estimate of an
unknown fact, like the number of humans on earth in 2050.

Consider further on that the agents are willing to adjust their opinion
towards the opinions of others. Adjustment of a continuous opinion can be
well described by computing a weighted arithmetic mean of other agent's
opinions. A process of continuous opinion dynamics can be seen as
repeated averaging of opinions.

If the averaging weights are fixed (as in
\cite{DeGroot1974,Berger1981,Chatterjee1977}) then the process can be
mapped to $x(t+1) = Ax(t)$ with $x(0)$ being the $n$-dimensional vector
of opinions and $A$ being a row-stochastic matrix which represents the
averaging weights for each agent in a row. The central research question
was about conditions for reaching consensus.

Krause \cite{1997} invented a nonlinear bounded confidence model based on
this linear model in 1997. Agents give positive weights to other agents
only when they are close in opinion to their own opinion. So, agents may
change their weights dynamically. Analytical conditions for convergence
to consensus are only possible for very low number of agents
\cite{Krause2000}, so in the next step extensive computer simulations
have been done together with Hegselmann \cite{Hegselmann2002}. Then, the
model got a lot of attention and is now mostly referenced as the
Hegselmann-Krause model.

Independently, Deffuant and others invented a similar bounded confidence
model working on a project \cite{IMAGES2001, Deffuant2000}about improving
agri-environmental policies in 2000. Partly inspired by Axelrod
\cite{Axelrod1997} and particle physics they proposed a model of random
pairwise interaction, where agents compromise if their opinions differ
not too much.

Both models differ a lot in the detail (e.g. one is stochastic, one is
deterministic) but are on the other hand similar in spirit, because they
both make a `bounded confidence'-assumption for the agents. They can be
represented as special cases in a general model \cite{Urbig2004}. For
this general model it is possible to prove convergence to a limit opinion
configuration \cite{Lorenz2005}. But the proof does not use the bounded
confidence assumption and nothing has been said about the set of all
possible limit opinion configurations. This paper is to show the set of
fixed points for both models (Section \ref{sec:agent-based-bounded}).
Although the answer is quite plausible the proof for the
Hegselmann-Krause model is not trivial.

Further on, both models have been redefined as density-based models. The
idea goes back to Ben-Naim et al \cite{Ben-Naim2003} in 2002 for the
Deffuant-Weisbuch model and has been copied for the Hegselmann-Krause
model \cite{Fortunato2005b}. Both can be approximated as state-discrete
interactive Markov chains as done in
\cite{Lorenz2005a,Lorenz2006,LorenzPhd2007} inspired by
\cite{Conlisk1976}.  It has been seen in simulation that the
density-based models also converge to limit opinion formations. Both
types of limit formations are of the same heuristic type. But a proof of
convergence is lacking.

This paper is furthermore to show the set of fixed points for the
density-based models in their approxiamtion as interactive Markov chain
(Section \ref{sec:imc}). Again the answer is quite plausible but the
proof for the Hegselmann-Krause model is not trivial. The results for the
fixed points of the interactive Markov chains are the central result of
this paper.

In dynamical systems analysis it is natural to start finding the set of
fixed points. Fixed points are stable opinion configurations.  We guess
that the processes in the agent-based as well as density-based model
converge to one point in their set of fixed points. There is strong
evidence from simulation for this conjecture, but a proof is lacking for
the density-based models. As a starting point, we present a Lyapunov
function which ensures that the density-based dynamics of the
Deffuant-Weisbuch model can not have cycles.

The proofs for the fixed points of the interactive Markov chains rely on
defining the difference equation which serves as a sort of discrete
master equation which give gain and loss terms for each opinion class. 

\section{Agent-based bounded confidence
  models}\label{sec:agent-based-bounded}

Let us consider a set of $n\in\N$ agents which hold continuous opinions.
An opinion is a real number or respectively a vector of $d\in\N$ real
numbers. The \emph{opinion space} is thus $S\subset \R^d$. Usually $S$ is
compact and convex. The opinion of agent $i\in\n$ at time $t\in\N$ is
$x_i(t)\in S$, and the vector $x(t) \in S^n$ is the \emph{opinion
  profile} at time $t$.  Notice that $x(t) \in (\R^d)^n$ is a vector of
vectors for $d>1$.

\begin{figure}[htbp]
  \centering
  \includegraphics[width=0.3\textwidth]{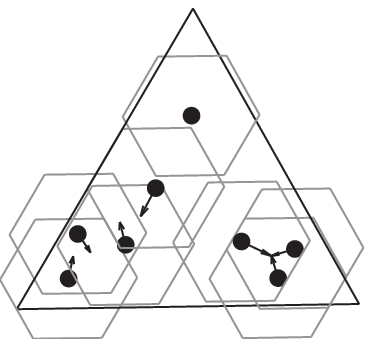}\quad
  \includegraphics[width=0.3\textwidth]{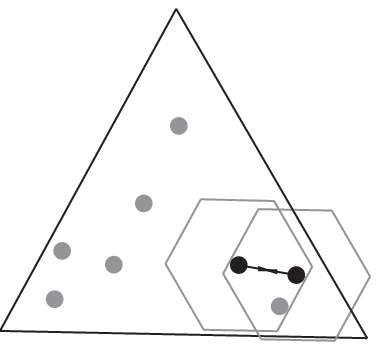}
  \caption{Visualisation for HK (left) and DW (right) dynamics in
    agent-based representation.}
  \label{fig:vis}
\end{figure}

Figure \ref{fig:vis} shows visualisations for the dynamics in one time
step of both processes in a two dimensional triangular opinion
space\footnote{This opinion space could represent a simplex where
  opinions are proposals for the allocation of a fixed amount of money to
  three projects. See \cite{Lorenz2006a,Lorenz2007} for simualtion
  results and the impact of the dimension in simplex opinion space.}

\subsection{Agent-based Hegselmann-Krause model}\label{mod:HK}

Let there be $n\in\N$ agents and an appropriate opinion space
$S\subset\R^d$.

Given an initial profile $x(0) \in S^n$, bound of confidence $\eps > 0$
and a norm $\norm{\cdot}$ we define the \emph{HK process}
$(x(t))_{t\in\N}$ recursively through
\begin{equation}
  x(t+1) = A(x(t),\eps) x(t),
\end{equation}
with $A(x,\eps)$ being the confidence matrix defined
\[
A_{ij}(x,\eps) := \left\{ \begin{array}{cl}
    \frac{1}{\#I_{\eps}(i,x)} \quad & \textrm{if } j\in I_{\eps}(i,x)   \\
    0 & \textrm{otherwise,}
  \end{array} \right. \\
\]
with $I_{\eps}(i,x) := \{j\in\n \,|\, \norm{x^i - x^j} \leq \eps \}$.

\subsection{Agent-based Deffuant-Weisbuch model}\label{mod:DW}

Let there be $n\in\N$ agents and an opinion space $S\subset\R^d$ convex.
Given an initial profile $x(0) \in S^n$, bound of confidence $\eps > 0$,
and a norm $\norm{\cdot}$ we define the \emph{DW process} as the random
process $(x(t))_{t\in\N}$ that chooses in each time step $t\in\N$ two
random agents $i,j$ which perform the action
\begin{eqnarray*}
  x^i(t+1) &=& \left\{
    \begin{array}{ll}
      \frac{1}{2}(x^j(t)+x^i(t)) & \hbox{if $\norm{x^i(t)-x^j(t)}\leq\eps$} \\
      x^i(t) & \hbox{otherwise.}
    \end{array}
  \right.\\
\end{eqnarray*}
The same for $x^j(t+1)$ with $i$ and $j$ interchanged.

\subsection{The set of fixed points in agent-based
  models}\label{sec:set-fixed-points}

We call $x^\ast\in(\R^d)^n$ a \emph{fixed point of the HK model} if
$A(x^\ast,\eps)x^\ast = x^\ast$. We call $x^\ast\in(\R^d)^n$ a
\emph{fixed point of the DW model} if for all choices $i,j\in\n$ the
profile $x^\ast$ does not change if agents $i$ and $j$ communicate.
Further, $F^{\HK}\subset (\R^d)^n$ and $F^{\DW}\subset (\R^d)^n$ are the
sets of fixed points of the corresponding models.

In the following we describe these sets and show that they are equal. The
proof for the DW model is trivial while the proof for the HK model needs
a little bit of care. It relies on the finiteness of the number of
agents.

The following lemma will be helpful. Beforehand we define for an opinion
profile $x\in(\R^d)^n$ and two agents $i,j\in\n$ as $H_{ij}\subset \R^d$
the hyperplane that is orthogonal to $x^j-x^i$ which goes through $x^j$,
and $H_{ij}^+\subset\R^d$ is the closed half-space defined by $H_{ij}$
which does not contain $x^i$.

\begin{lemma}\label{lem:fixABMHK}
  Let $x^\ast\in(\R^d)^n$ be a fixed point of the homogeneous HK model
  with bound of confidence $\eps>0$.  Let there be $i,j\in\n$ with
  $x^i\neq x^j$ such that $i\in I_\eps(j,x)$ and let there be $k\in\n$
  such that $x^k\not\in H_{ij}^+$. Then there exists $m\in I_\eps(j,x)$
  different from $i,j,k$ such that $x^m\in H_{kj}^+$.
\end{lemma}

\begin{proof}
  We abbreviate $x:=x^\ast$. Due to $x$ being a fixed point it must hold
  that
  \[x^j = \frac{1}{\#I_{\eps}(j,x)}\sum_{s\in I_{\eps}(j,x)}x^s.\] So,
  $x^j$ is the barycenter of all the opinions in of agents in the
  confidence set $I_{\eps}(j,x)$. By definition $i\in I_{\eps}(j,x)$. Due
  to the fact that $k\not\in H_{ij}^+$ and that $H_{ij}^+$ is closed, the
  angle between $x^i-x^j$ and $x^k-x^j$ is less then $\frac{\pi}{2}$ and
  thus $i\not\in H_{kj}^+$. (See Figure \ref{fig:visH} for a
  visualisation.)

\begin{figure}[htbp]
  \centering
  \includegraphics[width=0.3\textwidth]{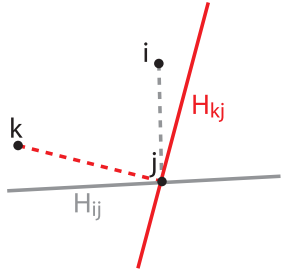}
  \caption{Visualisation of $H_{ij}^+$ and $H_{kj}^+$.}
  \label{fig:visH}
\end{figure}

There must be at least one more agent in $I_{\eps}(j,x)$ besides $i$ and
$j$, because otherwise $x^j$ is not the barycenter of $x^i$ and $x^j$. If
all these other agents were not in $H_{jk}^+$ then $x^j$ would be an
extreme point\footnote{See Rockafellar \cite{Rockafellar1970} for convex
  analysis.} of the convex hull of the opinions of the agents in
$I_{\eps}(j,x)$. Thus, there must be $m\in I_\eps(j,x)$ such that $x^m\in
H_{jk}^+$ and $m\neq j$.
\end{proof}

\begin{proposition}\label{prop:fixABM}
  Let $\eps>0$ be a bound of confidence which defines the homogeneous HK
  model and the DW model on the opinion space $(\R^d)^n$. It holds that
  \begin{equation}\label{eq:fixABM}
    F^{\HK}=F^{\DW}=\{x\in(\R^d)^n \,|\, \forall i,j\in\n : \norm{x^i - x^j}_p>\eps \hbox{ or } x^i = x^j\}.
  \end{equation}
\end{proposition}

\begin{proof}
  If $x$ is in the set as described in \eqref{eq:fixABM} each two agents
  either reached consensus or are too far away from each other to
  interact. Thus, $x$ is a fixed point in the DW and in the HK model.

  Let $x$ be not in the set as described in \eqref{eq:fixABM} then there
  are $i,j\in\n$ such that $\norm{x^i-x^j}_p\leq\eps$ and $x^i\neq x^j$.

  Then $x$ can not be a fixed point of the DW model, because if $i,j$ are
  chosen as communication partners both agents will move towards each
  other.

  It remains to show that $x$ cannot be a fixed point of the HK model. We
  assume that $x$ is a fixed point of the HK model and derive a
  contradiction.

  Due to Lemma \ref{lem:fixABMHK} there exists $m_0\in I_\eps(j,x)$ with
  $x^{m_0} \in H_{ij}^+$ with \[\norm{x^i-x^{m_0}}_2 > \norm{x^i-x^j}_2\]
  (we set $k$ in the lemma equal to $i$). Now we apply the lemma again
  for $j \in I_\eps(m_0,x)$. Then obviously $i\not\in H_{jm_0}^+$, and
  thus there is $m_1\in H_{im_0}^+$ such that $\norm{x^i-x^{m_1}}_2 >
  \norm{x^i-x^{m_0}}_2$.  We can conclude like this to derive a sequence
  of agents $m_0,m_1,m_2,\dots$ such that $\norm{x^i-x^{m_0}}_2 <
  \norm{x^i-x^{m_1}}_2 < \norm{x^i-x^{m_2}}_2 < \dots$. This is a
  contradiction to the finiteness of the number of agents.
\end{proof}

\begin{figure}[htbp]
  \centering
  \includegraphics[scale=0.8]{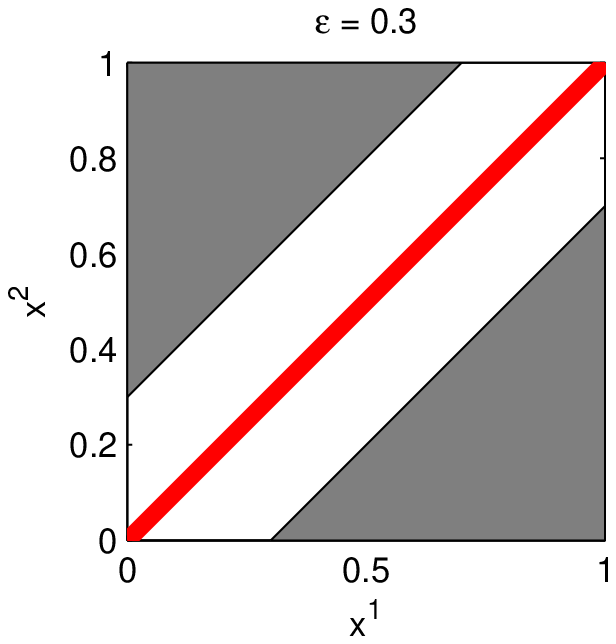}
  \includegraphics[scale=0.8]{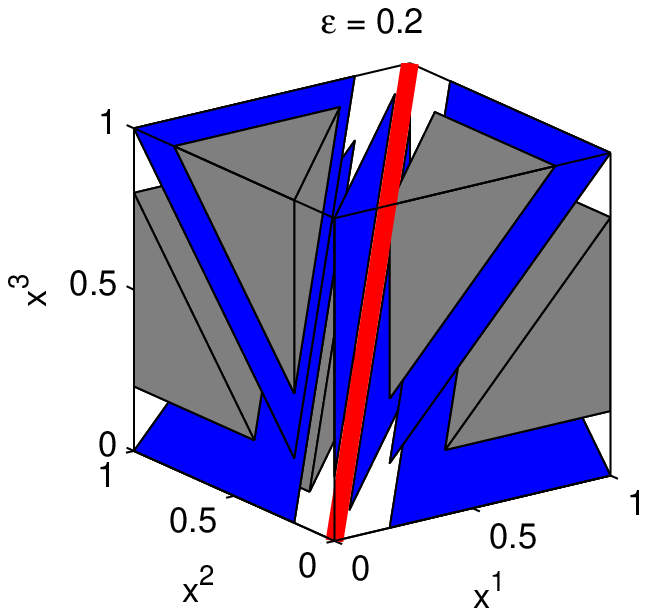}
  \caption{The set of fixed points $F^{\HK}$ and $F^{\DW}$ for the
    opinion space $[0,1]\subset\R$ (so $d=1$), $n=2,3$ and $\eps=0.3$.
    The red line represents all consensus points.  The blue patches all
    points where two agents found consensus, while the other is far
    enough away. The gray regions represent all fixed points where each
    agent has an individual opinion. The 'invisible' space are thus all
    points where dynamics happen. }
  \label{fig:fixABM}
\end{figure}

Figure \ref{fig:fixABM} gives impressions how the set of fixed points
$F^{\HK}$ and $F^{\DW}$ looks for the opinion space $[0,1]\subset\R$ (so
$d=1$), $n=2,3$ and $\eps=0.3$.

For higher $n$ (but still $d=1$) one can imagine this set like: Take the
whole state space $\R^n$ and remove successively points. First, take all
subspaces where numbers in two dimensions must be equal and remove the
closed $\eps$-region around this subspaces from the whole space but keep
the subspaces itself. Then take from every of these subspaces all
subspaces where either a third number must be equal to the former two, or
two other numbers must be equal and remove their $\eps$-region but keep
the subspaces them self. Continuing like this spans a lattice of
subspaces which is of the same kind as the lattice of partitions of the
set $\{1,2,\dots,n\}$. The number of subspaces to treat is much bigger
than $n$ it is determined by the Bell numbers.

\section{Density-based bounded confidence models}\label{sec:imc}

In the following we reformulate the Hegselmann-Krause model and
Deffuant-Weisbuch model for a one-dimensional opinion space as
density-based models with the same heuristics as in the agent-based
model. We approximate density-based dynamics as interactive Markov chains
as first outlined in \cite{Lorenz2005a}.

Instead of concrete agents and their opinions we define the state of the
system as a density function on the opinion space which evolves in time.
As a simplification we only regard a one-dimensional interval as opinion
space and discretise it into $n$ subintervals which serve as
\emph{opinion classes}. So, we switch from $n$ agents with opinions in
the opinion space to an idealized infinite population, which is divided
to the opinion classes $\n = \{1,\dots,n\}$.

Class $i$ contains a fraction of the total population $p_i$. For
convenience we define $p_i = 0$ for all $i\notin\n$. A vector $p(t) \in
\R^n$ represents the \emph{opinion distribution} at time $t\in\N$.
Naturally, the fractions in the classes should sum up to one.  So, the
state space in a density-based model is a simplex. We define
$\triangle^{n-1} = \{p\in\R^n_{\geq 0} \,|\, \sum_{i=1}^n p_i = 1\}$. One
should think of an opinion distribution as a row vector.

If we define transition probabilities from one class to another we can
represent the opinion dynamics process as an interactive Markov chain
with transition matrix $B(p(t))$. It is called `interactive' because the
transition matrix depends on the actual state of the system.

Let $\n$ be a set of opinion classes and
$p(0)\in\triangle^{n-1}\subset\R^n$ be an initial opinion distribution.
A \emph{density-based process} is defined as an interactive Markov chain
\begin{equation}
  p(t+1)=p(t)B(p(t))\label{eq:imc}
\end{equation}
with the explicit definition of the transition matrix function. In the
following we give $B^{\HK}(p(t))$ for the HK model and $B^{\DW}(p(t))$
for the DW model.

\subsection{Density-based Hegselmann-Krause transition
  matrix}\label{sec:dens-based-hegs}

We need some preliminary definitions to define the transition matrix for
the interactive Markov chain with communication of repeated meetings like
in the Hegselmann-Krause model.

Let $I=\{i,\stackrel{+1}{\dots},j\}\subset\n$ be a discrete interval and
$p\in \triangle^{n-1}$ be an opinion distribution. We call
\[
M_I^0(p) := \sum_{k\in I} p_k \textrm{ the \emph{$I$-mass} (or 0th
  moment) of $p$,}
\]
\[
M_I^1(p) := \sum_{k\in I} kp_k \textrm{ the \emph{first $I$-moment} of
  $p$ and}
\]
\[
M_I^\mathrm{bary}(p) := \left\{%
  \begin{array}{ll}
    \frac{M_I^1(p)}{M_I^0(p)}, & \hbox{if $p_I \neq 0$,} \\
    \frac{\max I + \min I}{2}, & \hbox{if $p_I = 0$.} \\
  \end{array}%
\right. \textrm{ the \emph{$I$-barycenter} of $p$.}
\]

Let $p\in \triangle^{n-1}$ be an opinion distribution and $\epsilon\in\N$
be a discrete bound of confidence. For $i\in\n$ we abbreviate the
\emph{$\epsilon$-local mean} as
\[
M_i :=
M^\mathrm{bary}_{\{i-\epsilon,\stackrel{+1}{\dots},i+\epsilon\}}(p)
\]
We define the \emph{HK transition matrix} as
\[
B^\HK_{ij}(p,\epsilon) := \left\{%
  \begin{array}{ll}
    1 & \hbox{if $j = M_i$,} \\
    \lceil M_i\rceil - M_i & \hbox{if $j = \lfloor M_i \rfloor$, $j\neq M_i$,} \\
    M_i - \lfloor M_i\rfloor & \hbox{if $j = \lceil M_i \rceil$, $j\neq M_i$,}  \\
    0 & \hbox{otherwise.} \\
  \end{array}%
\right.
\]

\begin{figure}[htbp]
  \centering
  \includegraphics[width=0.3\textwidth]{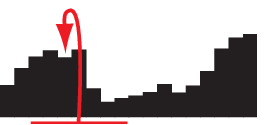}\hspace{2cm}
  \includegraphics[width=0.3\textwidth]{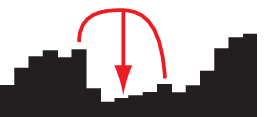}
  \caption{Visualisation of dynamics in density-based models. HK right,
    DW left}
  \label{fig:visdb}
\end{figure}

Each row of the transition matrix $B^\HK(p,\epsilon)$ contains only one
or two adjacent positive entries. The population with opinion $i$ goes
completely to the $\epsilon$-local mean opinion if this is an integer.
Otherwise they distribute to the two adjacent opinions.  The fraction
which goes to the lower (upper) opinion class depends on how close the
$\epsilon$-local mean lies to it. Thus, the heuristic of averaging all
opinions in a local area is represented. Figure \ref{fig:visdb} may give
a hint how dynamics work.

\subsection{Density-based Deffuant-Weisbuch transition
  matrix}\label{sec:dens-based-deff}

The \emph{Deffuant-Weisbuch transition matrix} for an opinion
distribution $p \in \triangle^{n-1}$, a discrete bound of confidence
$\epsilon \in \N$ is defined by
\[
B^\DW_{ij}(p,\epsilon,\mu) \left\{%
  \begin{array}{ll}
    \frac{\pi^i_{2j-i-1}}{2} + \pi^i_{2j-i} + \frac{\pi^i_{2j-i+1}}{2}, & \hbox{if $i\neq j$, } \\
    q_{i}, & \hbox{if $i=j$.} \\
  \end{array}%
\right.
\]
with $q_i = 1 - \sum_{j\neq i, j=1}^n B^\DW_{ij}(p,\epsilon,\mu)_{ij}$
and
\[\pi^i_m := \left\{%
  \begin{array}{ll}
    p_m, & \hbox{if $|i-m|\leq \epsilon$} \\
    0, & \hbox{otherwise} \\
  \end{array}%
\right.\]

Remember that we defined $p_i = 0$ for all $i\not\in \n$.

We briefly describe how the agent-based heuristics of the
Deffuant-Weisbuch model governs the transition matrix of the interactive
Markov chain. By the founding idea of the model an agent with opinion $i$
moves to the new opinion $j$ if he compromises with an agent with opinion
$i + 2(j-i) = 2j-i$. The probability to communicate with an agent with
opinion $2j-i$ is of course $p_{2j-i}$. Thus, the heuristic of random
pairwise interaction is represented. The terms
$\frac{\pi^i_{2j-i-1}}{2},\frac{\pi^i_{2j-i+1}}{2}$ stand for the case
when agents with opinion $i$ communicate with agents with opinion $j$,
but the distance $|i-j|$ is odd. In this case the population should go
with probability $\frac{1}{2}$ to one of the two possible opinion classes
$\lfloor\frac{i+j}{2}\rfloor,\lceil\frac{i+j}{2}\rceil$.  Figure
\ref{fig:visdb} may give a hint how dynamics work.

\subsection{The set of fixed points in density based
  models}\label{sec:set-fixed-points-1}

Here, we will prove that the set of fixed points of the interactive
Markov chains\begin{equation} p(t+1) =
  p(t)B^\mathrm{CR}(p(t),\epsilon)\label{eq:imcinsec}
\end{equation}
with DW and HK transition matrix is
\begin{equation}
  G^{\HK} = G^{\DW} = \{p \in \triangle^{n-1} \,|\, p^{}_k > 0 \Rightarrow p^{}_{m} = 0 \textrm{ for all }
  m\in \{ k-\epsilon, \stackrel{+1}{\dots}, k-2, k+2, \stackrel{+1}{\dots},
  k+\epsilon \} \cap \n \}.
  \label{eq:prop:fix}
\end{equation}

The structure of the set of fixed points is thus: all opinion classes
with positive mass lie in adjacent pairs or isolated. Pairs and isolated
classes must have a distance greater than $\epsilon$ to each other. In an
adjacent pair of classes in a fixed point there are no further
restrictions on the proportion of agents in the two classes. So, fixed
points lie in certain lines in the simplex $\triangle^{n-1}$.

Further on, we give a Lyapunov-function for the interactive Markov chain
with DW transition matrix which rules out cycles. Convergence to fixed
point remains as conjecture for the DW as well as for the HK transition
matrix.

For both interactive Markov chains it is useful to look at their
difference equation, because it can play the role of a discrete master
equation (see \cite{Gillespie1992}), which displays gain and loss terms
for the mass changes in one class at one time step.

Let
\begin{equation} \Delta p := pB^\mathrm{CR}(p,\epsilon) - p =
  p(B^\mathrm{CR}(p,\epsilon) - \unit),\label{eq:Delta}
\end{equation}
then the interactive Markov chain is a trajectory of the equation
\[
p(t+1) = p(t) + \Delta p(t) = p(t) + p(t)(B^\mathrm{CR}(p(t),\epsilon) -
\unit).
\]
An opinion distribution $p^\ast$ is a \emph{fixed point of the
  interactive Markov chain} \eqref{eq:imcinsec} if $p^\ast = p^\ast
B^\mathrm{CR}(p^\ast,\epsilon)$. Obviously, this is equivalent to $\Delta
p^\ast = 0$.

\subsubsection{The Deffuant-Weisbuch model}\label{sec:deff-weisb-model}

We take a look at the difference $\Delta p$ in detail. Simply calculating
equation (\ref{eq:Delta}) with $B^\mathrm{CR}(p,\epsilon) :=
B^\DW(p,\epsilon)$ leads to the the following explanatory difference
equation for all $k\in\n$.

\begin{equation}
  \Delta p_k = \underbrace{\sum_{\frac{i+j}{2}=k, 2\leq |i-j|\leq \epsilon}
    p_ip_j + \sum_{\frac{i+j}{2}=k\pm\frac{1}{2}, 2\leq |i-j|\leq \epsilon}
    \frac{1}{2}p_ip_j}_{\textrm{fraction joining state $k$}} -
  \underbrace{p_k\sum_{2\leq|j-k|\leq\epsilon} p_{j}}_{\textrm{fraction
      leaving state $k$}} \label{eq:i}
\end{equation}
(The first two sums go over all $(i,j)\in\n\times\n$, the third over
$j\in\n$, under restriction of the equations below.) This is analog to a
master equation in physics determining the fraction leaving a state and
the fraction joining a state, but discrete in state and time.

\begin{theorem} \label{thm:fixDW} An opinion distribution $p^{} \in
  \triangle^{n-1}\subset\R^n$ is a fixed point of the interactive Markov
  chain \eqref{eq:imcinsec} with DW transition matrix and discrete bound
  of confidence $\epsilon \in \N$ if and only if it holds for all
  $k\in\n$ that
  \begin{equation}
    p^{}_k > 0 \Rightarrow p^{}_{m} = 0 \textrm{ for all }
    m\in \{ k-\epsilon, \stackrel{+1}{\dots}, k-2, k+2, \stackrel{+1}{\dots}, k+\epsilon \} \cap \n.
    \label{eq:prop:fixwd}
  \end{equation}
\end{theorem}

\begin{proof} For the `if'-part let us assume that $p^{}$ is a fixed
  point and show that (\ref{eq:prop:fixwd}) holds. If $p$ is a fixed
  point it holds $\Delta p_k = 0$ for all $k\in\n$.

  Let $k\in\n$ be such that $p_k > 0$. For an indirect proof let us
  assume that there is $m_0\in \{ k-\epsilon, \dots, k-2, k+2,\dots,
  k+\epsilon \} \cap \n$ such that $p_{m_0} > 0$ and find a
  contradiction.

  We can conclude from $\Delta p_k = 0$ and equation (\ref{eq:i}) that it
  holds
  \[
  \underbrace{p_k}_{>0} \underbrace{\sum_{2\leq|j-k|\leq\epsilon}
    p_{j}}_{>0 \textrm{ because it contains $p_{m_0}$}} =
  \sum_{\frac{i+j}{2}=k, 2\leq |i-j|\leq \epsilon} p_ip_j +
  \sum_{\frac{i+j}{2}=k\pm\frac{1}{2}, 2\leq |i-j|\leq \epsilon}
  \frac{1}{2}p_ip_j
  \]

  Thus, on the right hand side one addend $p_{m_1}p_{n_1}$ must be
  positive. A careful look at the summation index sets will help us to
  conclude further. If we assume without loss of generality $m_1 < n_1$
  then we can conclude $m_1 < k$.

  We can conclude from $\Delta p^\ast_{m_1} = 0$ and equation
  (\ref{eq:i}) that
  \[
  \underbrace{p^\ast_{m_1}}_{>0} \underbrace{\sum_{2\leq|j-m_1|}
    p^\ast_{j}}_{>0 \textrm{ because it contains $p^\ast_{n_1}$}} =
  \sum_{\frac{i+j}{2}=m_1, 2\leq |i-j|\leq \epsilon} p_ip_j +
  \sum_{\frac{i+j}{2}=m_1\pm\frac{1}{2}, 2\leq |i-j|\leq \epsilon}
  \frac{1}{2}p_ip_j
  \]

  Thus, on the right hand side one addend $p_{m_2}p_{n_2}$ must be
  positive again and there is $m_2 < m_1 < k$.

  We conclude by induction until we reach an index $m_z < 1$ for which
  $p_{m_z}$ must be positive -- a contradiction.

  To prove the `only if'-part we assume that for all $k\in\n$ it holds
  (\ref{eq:prop:fixwd}). We have to check that $\Delta p_k=0$ in equation
  (\ref{eq:i}) for all $k\in\n$. We see that every addend in each
  equation is of the form $p_ip_j$ with $2\leq |j-l|\leq \epsilon$ and
  $\frac{i+j}{2} \in \{k, k\pm\frac{1}{2}\}$. From (\ref{eq:prop:fixwd})
  we know that in every case either $p_i$ or $p_j$ are zero.
\end{proof}

\begin{theorem}\label{thm:convDWIMC}
  For every $p(0) \in \triangle^{n-1}\subset\R^n$ the interactive DW
  Markov chain $(p(t))_{t\in\N_0}$ can not be periodic.
\end{theorem}

\begin{proof} We define a Lyapunov function $L: S_n \to \R$ which is
  continuous and strictly decreasing on $(p(t))_{t\in\N_0}$ for every
  initial distribution $p(0)$ as long as we do not reach a fixed point.
  Let

  \[L(p) := \sum_{i=1}^n 2^i p_i.\]

  Now we have to show that for every $p$ which is not a fixed point it
  holds that $L(p) > L(pB(p,\epsilon))$.

  Because of the linearity of $L$ we can transform the inequality such
  that we have to show
  \[0 > L(pB(p,\epsilon)-p) = L(\Delta(p)).\]

  Due to (\ref{eq:i}) it holds

\begin{eqnarray*}
  L(\Delta p) & = & \sum_{k\in\n}2^k \left( \sum_{\frac{i+j}{2}=k, 2\leq |i-j|\leq \epsilon}
    p_ip_j + \sum_{\frac{i+j}{2}=k\pm\frac{1}{2}, 2\leq |i-j|\leq \epsilon}
    \frac{1}{2}p_ip_j \right.\\
  & & \quad\quad\quad\quad\quad\quad\quad\quad\quad\quad\quad\quad\quad \left.  -p_k\sum_{2\leq|j-k|\leq\epsilon} p_{j} \right) \\
  &= &\sum_{2\leq |i-j|\leq \epsilon}(2^{\lfloor\frac{i+j}{2}\rfloor}+2^{\lceil\frac{i+j}{2}\rceil})p_ip_j +
  \sum_{2\leq |i-j|\leq \epsilon} (2^i+2^j)p_ip_j \\
  & = & \sum_{2\leq |i-j|\leq \epsilon}(2^{\lfloor\frac{i+j}{2}\rfloor}+2^{\lceil\frac{i+j}{2}\rceil} - 2^i - 2^j)p_ip_j
  \label{eq:liap}
\end{eqnarray*}

It holds $(2^{\lfloor\frac{i+j}{2}\rfloor}+2^{\lceil\frac{i+j}{2}\rceil}
- 2^i - 2^j) < 0$ for all $i,j$ with $|i-j| \geq 2$ and thus it holds
$L(\Delta p_i) <0$.
 
Due to the existence of the Lyapunov function it holds that
$(p(t))_{t\in\N_0}$ can not have cycles. Because if we consider that
there is a period $T\in\N$ such that $p(t) = p(t+T)$ then the sum
$\sum_{s=t}^{t+T-1}L(\Delta p(s))$ would be negative, but on the other
hand it also holds
\[
\sum_{s=t}^{t+T}L(\Delta
p(t))=\sum_{s=t}^{t+T}L(p(s+1)-p(s))=\sum_{s=t}^{t+T}L(p(s+1))-L(p(s))=0.
\]
Thus there is a contradiction to a periodic solution.
\end{proof}

If one would define a Lyapunov function which is zero on every fixed
point one might prove convergence to a fixed point.

\medskip

{\bf Conjecture.} For every $p(0) \in \triangle^{n-1}\subset\R^n$ the
interactive DW Markov chain $(p(t))_{t\in\N_0}$ converges to a fixed
point.

\medskip

There is evidence from simulation for this conjecture
\cite{LorenzPhd2007}.

\subsubsection{The Hegselmann-Krause model}\label{sec:hegs-krause-model}

Here we show that the fixed points of the interactive Markov chain
\eqref{eq:imc} with HK transition matrix are the same as for the DW transition matrix.

We start with a lemma on the $I$-barycenters.

\begin{lemma}
  Let $p\in \triangle^{n-1}\subset R^n$ be an opinion distribution and
  discrete intervals $I_0 = \{i_0,\stackrel{+1}{\dots},j_0\} \subset \n$
  and $I_1 = \{i_1,\stackrel{+1}{\dots},j_1\} \subset \n$. It holds
  \begin{enumerate}
  \item $i_0 \leq i_1$ and $j_0 \leq j_1 \Longrightarrow
    M^\mathrm{bary}_{I_0}(p) \leq M^\mathrm{bary}_{I_1}(p),$
    \label{prop:localmean:a}
  \item if $i_0 \leq i_1$ and $j_0 \leq j_1$
    \[ M^\mathrm{bary}_{I_0}(p) < M^\mathrm{bary}_{I_1}(p)
    \Longleftrightarrow \exists m \in (I_0 \cup I_1)\setminus (I_0 \cap
    I_1) \textrm{ with } p_m > 0.\] \label{prop:localmean:b}
  \end{enumerate} \label{lem:localmean}
\end{lemma}

\begin{proof}
  In a first step we assume $p_{I_0} \neq 0$ and $p_{I_1} \neq 0$ Thus
  there is $m_0\in I_0$ with $p_{m_0}>0$ and one $m_1\in I_1$ with
  $p_{m_1}>0$ thus the following equation is well defined:
  \begin{eqnarray*}
    M^\mathrm{bary}_{I_0}(p) &=& \frac{M^1_{I_0}(p)}{M^0_{I_0}(p)} =
    \frac{M^1_{I_0}(p)M^0_{I_1}(p)}{M^0_{I_0}(p)M^1_{I_1}(p)}M^\mathrm{bary}_{I_1}(p) \\
    &=& \frac{\sum\limits_{(m_0,m_1) \in I_0\times I_1} m_0
      p_{m_0}p_{m_1}}{\sum\limits_{(m_0,m_1) \in I_0\times I_1} m_1
      p_{m_0}p_{m_1}}M^\mathrm{bary}_{I_1}(p) \label{proof:localmean:1}
  \end{eqnarray*}

  To prove (\ref{prop:localmean:a}) we have to show that the fraction in
  equation (\ref{proof:localmean:1}) is less or equal than one.

  We compare the summands in the numerator and the denominator. If
  $m_0,m_1 \in I_0 \cap I_1$ then the summands $m_0p_{m_0}p_{m_1}$ and
  $m_1p_{m_0}p_{m_1}$ appear in both. In all other combination of indices
  it holds either $(m_0,m_1) \in (I_0\setminus I_1) \times I_1$ or
  $(m_0,m_1) \in I_0\times (I_1\setminus I_0)$. Due to $i_0 \leq i_1$ and
  $j_0 \leq j_1$ it holds $m_0 < m_1$ and thus the numerator is less or
  equal to the denominator and the fraction is less or equal to one.

  To prove (\ref{prop:localmean:b}) we have to show that fraction in
  (\ref{proof:localmean:1}) is strictly less then one. This holds if
  there is a pair $(m_0,m_1) \in (I_0\setminus I_1) \times I_1$ or
  $(m_0,m_1) \in I_0\times (I_1\setminus I_0)$ for which $p_{m_0}>0$ and
  $p_{m_1}>0$. This is obviously the case due to the claim in
  (\ref{prop:localmean:b}) and the assumption $p_{I_0} \neq 0$ and
  $p_{I_1} \neq 0$.

  At least we have to check the case, where $p_{I_0} = 0$ or $p_{I_1} =
  0$. The same steps as in Equation (\ref{proof:localmean:1}) lead with
  the definition of the local mean to the equations
  \begin{eqnarray*}
    p_{I_0} \neq 0, p_{I_1} = 0 & \Rightarrow & M_{I_0}^\mathrm{bc}(p)
    = \frac{2\sum_{m\in I_0\setminus I_1} m p_m}{(j_1 + i_1)\sum_{m\in
        I_0\setminus I_1} p_m}
    M_{I_1}^\mathrm{bc}(p) \\
    p_{I_0} = 0, p_{I_1} \neq 0 & \Rightarrow & M_{I_0}^\mathrm{bc}(p)
    = \frac{(j_0 + i_0)\sum_{m\in I_1\setminus I_0} p_m}{2\sum_{m\in
        I_1\setminus I_0} m p_m}
    M_{I_1}^\mathrm{bc}(p) \\
    p_{I_0} = 0, p_{I_1} = 0 & \Rightarrow & M_{I_0}^\mathrm{bc}(p) =
    \frac{j_0 + i_0}{j_1 + i_1} M_{I_1}^\mathrm{bc}(p)
  \end{eqnarray*}
  (We can choose the summation index sets $I_0\setminus I_1$ instead of
  $I_0$ in the upper equation, because all summands with indices out of
  $I_0 \cap I_1$ are obviously zero. Analog for the middle equation.) For
  all three equations we can conclude like above to get
  \ref{prop:localmean:a} and \ref{prop:localmean:b}.
\end{proof}

For an opinion distribution $p\in \triangle^{n-1}\subset\R^n$ and a
discrete bound of confidence $\epsilon\in\n$ we recall the abbreviation
$M_i :=
M^\mathrm{bary}_{\{i-\epsilon,\stackrel{+1}{\dots},i+\epsilon\}}(p)$.
Due to Lemma \ref{lem:localmean} it holds
\begin{equation}
  M_1 \leq M_2 \leq \dots \leq M_n.\label{eq:localmeanineq}
\end{equation}

Analog to the former subsection we reformulate \eqref{eq:Delta}, which
leads to the following explanatory difference equation for all $k\in\n$
(again in analogy to a master equation).

\begin{equation}
  \Delta p_k = \underbrace{\sum_{j\in I_k^{\lceil\cdot\rceil}} (M_j
    - \lfloor M_j \rfloor) p_j + \sum_{j\in I_k^{}} p_j +
    \sum_{j\in I_i^{\lfloor\cdot\rfloor}} (\lceil M_j \rceil - M_j)
    p_j}_{\textrm{fraction joining $k$}} 
  \underbrace{ - p_k}_\textrm{fraction leaving $k$}
  \label{eq:ihk}
\end{equation}

with
\[
I_k^{\lceil\cdot\rceil} := \{j\in\n\,|\, M_j \neq k = \lceil M_j \rceil
\textrm{ and } p_k>0\},
\]
\[
I_k^\ast := \{j\in\n\,|\, k = M_j \textrm{ and } p_k>0\}\textrm{ and }
\]
\[
I_k^{\lfloor\cdot\rfloor} := \{j\in\n\,|\, M_j \neq k = \lfloor M_j
\rfloor \textrm{ and } p_jk>0\}.
\]

It is easy to see with \eqref{eq:localmeanineq} that the sets
$I_i^{\lceil\cdot\rceil}, I_i^\ast$ and $I_i^{\lfloor\cdot\rfloor}$ are
all discrete intervals, that they are pairwise disjoint and that their
union
\[
I_i := I_i^{\lceil\cdot\rceil} \cup I_i^\ast \cup
I_i^{\lfloor\cdot\rfloor}
\]
is a discrete interval, too. We know also that the coefficients $(M_j -
\lfloor M_j \rfloor)$ and $(\lceil M_j \rceil - M_j)$ in (\ref{eq:ihk})
are always positive and strictly less than one by definition.

The following proposition shows that an opinion class with positive mass
has a local barycenter which is less than one class away and that the
adjacent class has positive mass too and a local barycenter between the
two classes.

\begin{proposition}\label{prop:imc:hk:prop2}
  Let $p \in \triangle^{n-1}\subset\R^n$ be a fixed point of the
  interactive HK Markov chain \eqref{eq:imcinsec} with HK transition
  matrix and let $p_i>0$ then it holds
  \begin{eqnarray*}
    \textrm{either} & & M_i = i \textrm{ and } I_i = \{i\}\\
    \textrm{or} & & i<M_i\leq M_{i+1}<i+1, p_{i+1} > 0 \textrm{ and
    } I_i = \{i,i+1\} = I_{i+1}, \\
    \textrm{or} & & i-1<M_{i-1}\leq M_{i}<i, p_{i-1} > 0 \textrm{ and
    } I_i = \{i,i-1\} = I_{i-1} 
  \end{eqnarray*}
\end{proposition}

\begin{proof}
  We define $p = [p_1 \dots p_n]$.

  In a first step we will show that $i-1 < M_i < i+1$. Let us assume for
  an indirect proof that $M_i \geq i+1$.

  The fact that $p$ is a fixed point implies $\Delta p = 0$ and thus we
  can derive from Equation (\ref{eq:ihk}) that
  \begin{equation}
    p_i = \sum_{j\in I_i^{\lceil\cdot\rceil}} (M_j - \lfloor M_j
    \rfloor) p_j + \sum_{j\in I_i^{\ast}} p_j + \sum_{j\in
      I_i^{\lfloor\cdot\rfloor}} (\lceil M_j \rceil - M_j) p_j
    \label{eq:prop:hk1:1}
  \end{equation}

  Due to $M_i \geq i+1$ it holds that $i \not\in I_i$ (the union of all
  index sets) and due to Lemma \ref{lem:localmean} it holds for $j\in
  I_i$ that $j\leq i-1$. Let $i_1 := \max I_i$. Thus it is clear that
  $i_1 < i, p_{i_1} > 0$ and $i_1 < M_{i_1}$.

  We conclude further with Equation (\ref{eq:ihk}) that
  \begin{equation}
    p_{i_1} = \sum_{j\in I_{i_1}^{\lceil\cdot\rceil}} (M_j - \lfloor
    M_j \rfloor) p_j + \sum_{j\in I_{i_1}^{\ast}} p_j + \sum_{j\in
      I_{i_1}^{\lfloor\cdot\rfloor}} (\lceil M_j \rceil - M_j) p_j
    \label{eq:prop:hk1:2}
  \end{equation}
  It may $i_1 \in I_{i_1}^{\lceil\cdot\rceil}$ but it holds $\max I_{i_1}
  \leq i_1$ and due to $(M_j - \lfloor M_j \rfloor) < 1$ it holds that
  there must exist $i_2 := \max I_{i_1}\setminus\{i_1\}$ with $p_{i_2} >
  0$ and $i_2 < M_{i_2}$.

  We derive by induction further on the existence of a decreasing chain
  of indices $i > i_1 > i_2 > \dots$ with $p_{i} >0,p_{i_1} >0,p_{i_2}
  >0,\dots$. Thus there must be $z < 1$ with $p_z > 0$, a contradiction,
  thus $M_i < i+1$.

  If we assume $M_i\leq i-1$ we can derive analog that there must be
  $z>n$ with $p_z > 0$. Thus we know $i-1 < M_i < i+1$.

  In the second step we show $ M_i > i \Rightarrow M_{i+1} < i+1, p_{i+1}
  > 0$. It is clear by Lemma \ref{lem:localmean} that $M_{i+1} \geq M_i$,
  lets assume $M_{i+1} \geq i+1$. Then we find (looking at Equation
  (\ref{eq:prop:hk1:1})) that $i \in I_i^{\lfloor\cdot\rfloor}$ and $i+1
  \not\in I_i$ thus we can conclude in the same way as after Equation
  (\ref{eq:prop:hk1:2}) that there exist $z<1$ with $p_z>0$. Thus it
  follows by this contradiction that $M_{i+1} < i+1$. Analog we derive $
  M_i < i \Rightarrow M_{i-1} > i-1$

  In the third step we show that $M_i > i$ implies $I_i = \{i,i+1\} =
  I_{i+1}$ and $p_{i+1}>0$. From equation $\Delta p = 0$ and Equation
  (\ref{eq:ihk}) we can derive the two equations
  \begin{eqnarray*}
    p_i & = & (\lceil M_i \rceil - M_i)p_i + (\lceil M_{i+1} \rceil -
    M_{i+1})p_{i+1} + \sum_{j\in
      I_i\setminus\{i,i+1\}}\textrm{positive terms} \\
    p_{i+1} & = & (M_i - \lfloor M_i \rfloor)p_i + (M_{i+1} - \lfloor M_{i+1} \rfloor)p_{i+1} + \sum_{j\in
      I_{i+1}\setminus\{i,i+1\}}\textrm{positive terms}
  \end{eqnarray*}
  If we add both equations we get by calculation
  \[
  0 = \sum_{j\in I_i\setminus\{i,i+1\}}\textrm{positive terms} =
  \sum_{j\in I_{i+1}\setminus\{i,i+1\}}\textrm{positive terms}
  \]
  and thus $I_i\setminus\{i,i+1\}$ and $I_{i+1}\setminus\{i,i+1\}$ must
  be empty. And due $i+1 \in I_i$ it holds $p_{i+1}>0$.

  Analog, we prove that $M_i < i$ implies $I_i = \{i-1,i\} = I_{i-1}$ and
  $p_{i-1}>0$.
\end{proof}

So, for the fixed point $p$ and $p_i>0$ we know that either $I_i = \{i\}$
or $I_i=\{i,i+1\}$ with $p_{i+1}>0$ or $I_i=\{i-1,i\}$ with $p_{i-1}>0$.
We define two new discrete intervals
\[I_i^{-\epsilon} := \{(\min I_i)-\epsilon,\stackrel{+1}{\dots},(\min
I_i) - 1\},\]
\[I_i^{+\epsilon} := \{(\max I_i) + 1,\stackrel{+1}{\dots},(\max
I_i)+\epsilon\}.\] The discrete interval $I_i^{-\epsilon} \cup I_i \cup
I_i^{+\epsilon}$ is the interval which contains all the classes where the
imaginary agents in the classes of $I_i$ interact with.  The next
proposition shows that the class(es) in $I_i^{-\epsilon}$ and
$I_i^{+\epsilon}$ can only both contain mass or both contain no mass.

\begin{proposition}\label{prop:imc:hk:prop3}
  Let $p \in \triangle^{n-1}\subset\R^n$ be a fixed point of the
  interactive HK Markov chain \eqref{eq:imcinsec} with HK transition
  matrix and let $p_i>0$ then it holds
  \[
  p_{I_i^{-\epsilon}} = 0 \Leftrightarrow p_{I_i^{+\epsilon}} = 0.
  \]
\end{proposition}

\begin{proof}
  First we consider $I_i = \{i,i+1\}$. Thus, due to Proposition
  \ref{prop:imc:hk:prop2} it holds $i < M_i \leq M_{i+1} < i+1$. It holds
  $\Delta p = 0$ because $p$ is a fixed point. From \eqref{eq:ihk} we can
  thus derive
  \[
  p_i = (\lceil M_i \rceil - M_i)p_i + (\lceil M_{i+1} \rceil - M_{i+1})
  p_{i+1}
  \]
  With $\lceil M_i \rceil = \lceil M_{i+1} \rceil = i+1$ it follows
  \[
  p_i = ((i+1) - M_i)p_i + ((i+1) - M_{i+1})p_{i+1}.
  \]
  This can be transformed to
  \begin{equation}
    M_i p_i + M_{i+1} p_{i+1} = ip_i + (i+1)p_{i+1}\label{eq:imc:hk:prop3}
  \end{equation}
  Now, we assume for an indirect proof that $p_{I_i^{-\epsilon}} = 0$ and
  $p_{I_i^{+\epsilon}} \neq 0$ and derive a contradiction.  Due to this
  assumption it holds $M_i =
  M^\mathrm{bary}_{\{i,\stackrel{+1}{\dots},i+\epsilon\}}$ and $M_{i+1} =
  M^\mathrm{bary}_{\{i,\stackrel{+1}{\dots},i+1+\epsilon\}}$. Then it
  follows from lemma \ref{lem:localmean} that
  $M^\mathrm{bary}_{\{i,i+1\}} < M_{i+1}$ and
  $M^\mathrm{bary}_{\{i,i+1\}} \leq M_i$.  Now, we conclude from
  \eqref{eq:imc:hk:prop3} that
  \[M^\mathrm{bary}_{\{i,i+1\}} p_i + M^\mathrm{bary}_{\{i,i+1\}} p_{i+1}
  < ip_i + (i+1)p_{i+1}.\] Both sides divided by the positive term $(p_i
  + p_{i+1})$ delivers
  \[M^\mathrm{bary}_{\{i,i+1\}} < \frac{ip_i + (i+1)p_{i+1}}{p_i +
    p_{i+1}} = M^\mathrm{bary}_{\{i,i+1\}}.\]

  A similar contradiction can be derived for the assumption
  $p_{I_i^{-\epsilon}} \neq 0$ and $p_{I_i^{+\epsilon}} = 0$. This proves
  $p_{I_i^{-\epsilon}} = 0 \Leftrightarrow p_{I_i^{+\epsilon}} = 0$.

  For $I_i = \{i-1,i\}$ arguments are the same after renumbering $i \to
  i-1$.

  For $I_i=\{i\}$ it holds $M_i=i$. Again, we assume for an indirect
  proof that $p_{I_i^{-\epsilon}} = 0$ and $p_{I_i^{+\epsilon}} > 0$ and
  derive a contradiction:
  \[M_i = M^\mathrm{bary}_{\{i,\stackrel{+1}{\dots},i+\epsilon\}} <
  M^\mathrm{bary}_{\{i\}} = i = M_i.\]
\end{proof}

Now, we show that the set of fixed points of the interactive Markov chain
with HK transition matrix is the same as for the DW transition matrix.

\begin{theorem} \label{thm:fixHK} An opinion distribution $p^{} \in
  \triangle^{n-1}\subset\R^n$ is a fixed point of the interactive Markov
  chain \eqref{eq:imcinsec} with HK transition matrix and discrete bound
  of confidence $\epsilon \in \N$ if and only if it holds for all
  $k\in\n$ that
  \begin{equation}
    p^{}_k > 0 \Rightarrow p^{}_{m} = 0 \textrm{ for all }
    m\in \{ k-\epsilon, \dots, k-2, k+2, \dots, k+\epsilon \} \cap \n.
    \label{eq:prop:fixhk}
  \end{equation}
\end{theorem}

\begin{proof}
  For the `if'-part let us assume that $p$ is a fixed point and show that
  \eqref{eq:prop:fixhk} holds. For an indirect proof we assume that there
  are $i,j\in\n$ such that $i<j$, $2\leq|i-j|\leq\epsilon$ and
  $p_i,p_j>0$ and find a contradiction.

  From Proposition \ref{prop:imc:hk:prop2} we know that $I_i$ and $I_j$
  are disjoint.  From Proposition \ref{prop:imc:hk:prop3} we know that
  there must exist $m_0\in\N$ such that $m_0 < i$, $|i-m_0|\leq\epsilon$,
  $p_{m_0}>0$ and $I_{m_0}$ and $I_i$ are disjoint.  Comparing $m_0$ and
  $i$ we know with the same arguments that there must exist $m_1 \in \N$
  with $m_1< m_0$, $|m_0-m_1|\leq\epsilon$, $p_{m_1}>0$ and $I_{m_1}$ and
  $I_{m_0}$ are disjoint.  By induction we can construct a sequence of
  natural numbers $m_0 > m_1 > m_2 > \dots$ with
  $p_{m_0},p_{m_1},p_{m_2},\dots > 0$. Thus there must exist $z\in\N$
  such that $m_z<1$ and $p_{m_z}>0$, which is a contradiction.

  \medskip

  To prove the `only if'-part we assume that for all $k\in\n$ it holds
  (\ref{eq:prop:fixhk}). We have to check that $\Delta p_i=0$ in
  (\ref{eq:ihk}) for all $i\in\n$. We see that every addend in each
  equation is of the form $p_ip_j$ with $2\leq |j-l|\leq \epsilon$ and
  $\frac{i+j}{2} \in \{k, k\pm\frac{1}{2}\}$. From (\ref{eq:prop:fixhk})
  we know that in every case either $p_i$ or $p_j$ are zero.
\end{proof}

The convergence to a fixed point remains as a conjecture.

\medskip

{\bf Conjecture.} For every $p(0) \in \triangle^{n-1}\subset\R^n$ the
interactive HK Markov chain $(p(t))_{t\in\N_0}$ converges to a fixed
point.  Convergence occurs in finite time.

\medskip

There is strong evidence from simulation for the conjecture
\cite{LorenzPhd2007}.

\section{Conclusion}\label{sec:conclusion}

We characterised the set of fixed points for the agent-based DW and HK
model. They are identical. We did the same for their
corresponding density-based model versions (in the approximation of an
interactive Markov chain).

The proofs were not in every case trivial (especially in the
density-based HK model) although the set of fixed-points is quite
plausible on a first view. One reason for this is that there can be
arbitrary long covergence times in the HK model (for examples see
\cite{LorenzPhd2007}). 

Proofs of convergence for the interactive Markov
chains are still lacking, although there is strong evidence from
simulation for convergence to one point in the set of fixed
points. Further on, this
is an interesting type of set convergence. The processe 
processes show an interesting type of set-convergence. In contrast to
many other models these models have a huge amount of fixed points and
more over they are not isolated but appear in lines, planes and
hyperplanes. On the other hands in contrast to other types of
set-convergence the process always converges to one of these fixed points
and there are no limit cycles. 

A last question is about a class of models for which one can prove that
they have the presented sets of fixed points where both models appear as
special cases. Here, proofs for both models have been derived seperately,
although the models are similar in spirit. 

\bibliographystyle{plain}

\end{document}